\documentclass{iau}
\usepackage{graphicx}
\usepackage[english]{babel}
\usepackage{aas_macros}
\usepackage[pdfborder={0 0 0}]{hyperref}

\usepackage{natbib}
\title[Robust Foregrounds Removal] 
{Robust Foregrounds Removal\\for 21-cm Experiments}

\author[Florent Mertens et al.]   
{F. Mertens$^1$,
A. Ghosh$^{2,3}$
\and L.V.E. Koopmans$^1$
}

\affiliation{
$^1$Kapteyn Astronomical Institute, University of Groningen, \\
P. O. Box 800, 9700 AV Groningen, The Netherlands \\
[\affilskip]
$^2$Department of Physics and Astronomy, University of the Western Cape, \\
Robert Sobukwe Road, Bellville 7535, South Africa \\
[\affilskip]
$^3$Square Kilometre Array radio telescope (SKA) South Africa, \\
The Park, Park Road, Cape Town 7405, South Africa}

\pubyear{2018}
\volume{333}  
\jname{Peering towards Cosmic Dawn}
\editors{Vibor Jeli\'c \& Thijs van der Hulst, eds.}
\begin{document}

\maketitle

\begin{abstract}
Direct detection of the Epoch of Reionization via the redshifted 21-cm line will
have unprecedented implications on the study of structure formation in the early
Universe. To fulfill this promise current and future 21-cm experiments will need
to detect the weak 21-cm signal over foregrounds several order of
magnitude greater. This requires accurate modeling of the galactic and
extragalactic emission and of its contaminants due to instrument chromaticity,
ionosphere and imperfect calibration. To solve for this complex modeling, we
propose a new method based on Gaussian Process Regression (GPR)
which is able to cleanly separate the cosmological signal from most of the
foregrounds contaminants. We also propose a new imaging method based on a
maximum likelihood framework which solves for the interferometric equation
directly on the sphere. Using this method, chromatic effects causing the
so-called ``wedge" are effectively eliminated (i.e. deconvolved) in the
cylindrical ($k_{\perp}, k_{\parallel}$) power spectrum.

\keywords{methods:data analysis, statistical; techniques:interferometric;
cosmology: observations, early universe, large-scale structure of universe
}

\end{abstract}

\firstsection 
\section{Introduction}

Observations of the redshifted 21-cm signal from neutral hydrogen is a unique
probe of the early universe and can open the entire redshift window $z\sim6-30$
for astrophysical and cosmological studies, allowing us to directly study the
astrophysical processes occurring during the Epoch of Reionization (EoR) and the
Cosmic Dawn (CD). This exciting goal is challenged by the difficulty of
extracting the feeble 21-cm signal buried under astrophysical foregrounds orders
of magnitude brighter and contaminated by numerous instrumental systematics.

Several experiments are currently underway aiming at statistically detecting the
21-cm signal from the Epoch of Reionization (e.g. LOFAR, MWA, PAPER), already
achieving increasingly attractive upper limits on the 21-cm signal power
spectra~\citep{Patil17,Beardsley16,Ali15}, and paving the way for the second
generation experiments such as the SKA and HERA which will be capable of robust
power spectra characterization and for the first time directly image the large
scale neutral hydrogen structures from EoR and CD.

To fully exploit the sensitivity of these experiments, accurate removal of the
foregrounds is required. The radiation from our own Galaxy and other
extra-galactic sources are well-known to vary smoothly in frequency, and this
characteristic can be used to model and remove them~\citep{Jelic08,Chapman13}.
However, the interaction of the spectrally smooth foregrounds with the Earth’s
ionosphere and the observing instrument create additional ``mode-mixing''
foregrounds contaminants, which can mimic the 21-cm signal. Those are mainly due
to the rapid phase and sometime amplitude modifications of radio waves caused by
small-scale structures in the ionosphere~\citep[e.g.][]{Koopmans10}, the
inherent chromatic response of the instrument which creates chromatic side-lobe
noise~\citep[e.g.][]{Vedantham12,Thyagarajan15,Bharat17}, calibration
errors and mis-subtraction of sources due to imperfect sky modeling which also
introduce frequency structure to the otherwise smooth
foregrounds~\citep[e.g.][]{Patil16,Ewall17}.
Mitigating those sources of chromatic noise has proven to be extremely
difficult. The accuracy of the sky model used for calibration is limited by the
instrument confusion noise level and the precision of the beam response model of
our radio receivers. Unavoidably the observed signal will be
contaminated by mode-mixing.

To reduce the impact of this potential ‘show-stopper’, we have developed a
Maximum-Likelihood Power Spectra estimator using the spherical-wave visibility
equation (SpH ML) and a foreground removal method based on Gaussian Process
Regression (GPR).

\section{Spherical Harmonics Maximum-Likelihood Inversion}

 \begin{figure}
    \includegraphics{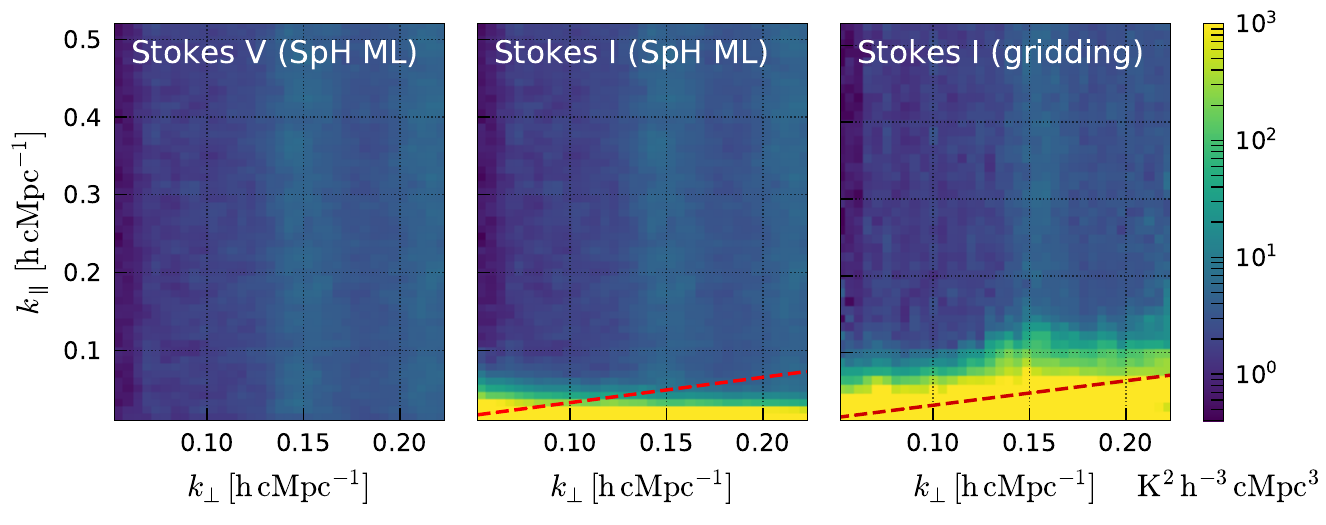}
    \caption{
Cylindrically averaged power spectra estimated from the simulated visibilities.
The central and right panel shows the power spectra of the observed signal
(foregrounds, noise and 21 cm signal; Stokes I) obtained from the spherical
harmonics ML inversion (central) and by gridding the simulated visibilities
using WSClean (right). The absence of structure inside the 10 Field of View
wedge line (red dashed line) in the power spectra estimated using the SpH ML
inversion demonstrate that the method effectively compute PSF-deconvolved
representation of the sky. The power spectra of the noise (Stokes V) is plotted
in the right panel.}
    \label{fig:ps2d_I+V_80sb}
 \end{figure}

The SpH ML is a new method to produce Point-Spread-Function (PSF) deconvolved
images of the sky from radio-interferometric observations. It solves for the
interferometric equation using Maximum-Likelihood inversion of the
spherical-wave visibility equation, formulated in a full sky setting and
including the primary beam and its side-lobes~\citep{Carozzi15,Ghosh17}:
\begin{equation}
\mathcal{V}=\sum_{lm}\tilde{v}_{\ell m}j_{\ell}(kr)Y_{\ell m}(\Omega_{k}).
\label{eq:VisibilitySphDecomp}
\end{equation}
The method is tested using simulated LOFAR-HBA full-sky observations which
include diffuse astrophysical foregrounds~\citep{Jelic08}, 21-cm signal
simulated using 21cmFast~\citep{Mesinger11} and noise level corresponding to 100
nights of 12 hr LOFAR integration time. From the simulated Stokes-I and Stokes-V
visibilities, we infer the recovered spherical harmonics using the SpH ML algorithm
and compute the angular power spectra. Figure~\ref{fig:ps2d_I+V_80sb} presents
the cylindrically averaged power spectra. We find the smooth diffuse foreground
in the Stokes I power spectrum mostly dominates at low $k_{\parallel}$, where
most of the foreground power is bound within $k_{\parallel} \le 0.05 \rm{h
cMpc^{-1}}$. The power drops by two to three orders of magnitudes in high
$k_{\parallel}$ regions, where the 21-cm signal plus noise signal is expected to
dominate. In the power spectra obtained from the same simulated visibilities but
using the more traditional method of gridding the visibilities in uv-space, a
wedge like structure is clearly visible (central panel of
Figure~\ref{fig:ps2d_I+V_80sb}), and is well known to be due to the frequency
dependence of the PSF~\citep{Vedantham12,Hazelton13}. This demonstrate that by
doing a ML fit to non-gridded visibility data sets at each frequency, we
effectively obtain a PSF-deconvolved estimates of the sky spherical harmonics
coefficients.

\section{Statistical 21-cm Signal Separation}

\begin{figure}
    \includegraphics[trim={2mm 2mm 0 0},clip]{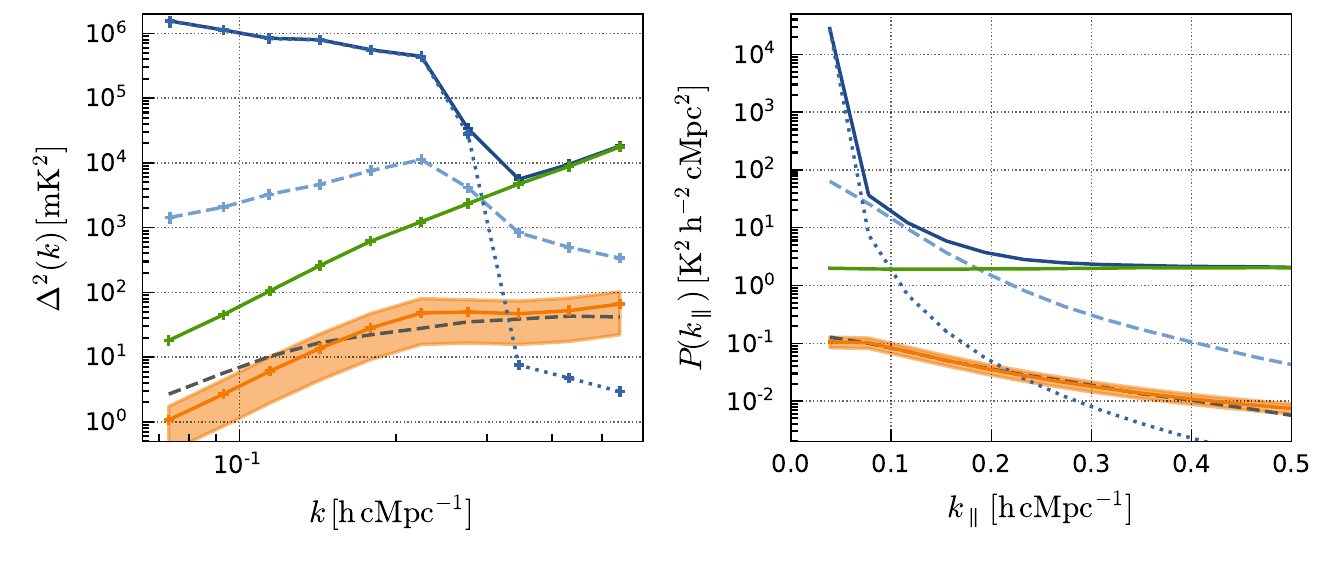}
    \caption{\label{fig:ps_reference} Detection of the EoR signal with the
    reference simulation. The top panel shows the spherically averaged power
    spectra. The central and bottom panels show the cylindrically averaged power
    spectra as a function of $k_{\perp}$ and $k_{\parallel}$ respectively. The
    simulated observed signal (dark blue) is composed of intrinsic astrophysical
    foregrounds (dotted dark blue), instrumental mode-mixing contaminants
    (dashed light blue), noise (green) and a simulated 21-cm signal (dashed
    gray). Using our GPR method to model and remove the foregrounds from the
    simulated cube, the 21-cm signal (orange) is well recovered with limited
    bias.}
\end{figure}

The source of mode-mixing contaminants are manifold, and we also need to account
for the remaining contaminants related to calibration error and sky-model
incompleteness. Ad-hoc modeling is not an option for most of them, and the
problem boils down to statistically separating this contaminants from the 21-cm
signal.

With this aim, we developed a foreground removal method using the technique of
Gaussian Process Regression (GPR)~\citep{Mertens17}. In this framework, the
different components of the observations, including the astrophysical
foregrounds, mode-mixing contaminants, and the 21-cm signal, are modeled with
Gaussian Process (GP). Formally, a GP is the joint distribution of a collection
of normally distributed random variables~\citep{Rasmussen05}. The covariance
matrix of this distribution is specified by a covariance function, which defines
the covariance between pairs of observations (e.g. at different frequencies).
The covariance function determines the structure that the GP will be able to
model, for example its smoothness. In GPR, we use GP as parameterized priors,
and the Bayesian likelihood of the model is estimated by conditioning this prior
to the observations. Standard optimization or MCMC methods can be used to
determine the parameters of the covariance functions.

This formulation ensures a relatively unbiased separation of their contribution
and accurate uncertainty estimation, even in very low signal to noise
observations. When applied to simulation datasets, equivalent to LOFAR-HBA 1200
hours of observations and based on its current assessment of noise and
systematic errors, we found that the method is capable of recovering well the
21-cm signal power spectrum (see Fig.~\ref{fig:ps_reference}).

\section{Discussions and Future Perspective}

In this paper we have introduced two new methods that aim at mitigating the
foregrounds contaminants including the mode-mixing.

We have shown that using an ML spherical harmonic power spectra estimator,  it
is possible to deconvolve the chromatic ``wedge" (caused by frequency-dependent
side-lobes) in the ($\rm{k}_{\perp}, \rm{k}_{\parallel}$) power spectrum space.

We also introduced a novel signal separation method which uses Gaussian
Process Regression (GPR) to model the various mixed components of the observed
signal, including the spectrally smooth sky, mode-mixing, and a 21-cm signal
model. 

The fundamental improvement of GPR resides in its complete statistical
description of all components contributing to the observed signal. In the
current implementation of GPR, we use generic covariance models for the 21-cm
signal and mode-mixing components. While this treatment may be sufficient for a
detection of the 21-cm signal and its characterization with LOFAR, an improved
model may be build for future experiments with e.g. the more sensitive SKA. The
mode-mixing model for example can be improved by integrating the $k_{\perp}$
dependency of the foreground wedge and folding into the model the analytic work
describing the effect on the signal of the instrumental chromaticity,
calibration errors and sky-model incompleteness. Exploiting the isotropic nature
of the 21-cm signal and its evolution at different redshift-bins will also
ensure a more sensitive and accurate modeling.

\bibliography{references}

\begin{thebibliography}{}
\makeatletter
\relax
\def\mn@urlcharsother{\let\do\@makeother \do\$\do\&\do\#\do\^\do\_\do\%\do\~}
\def\mn@doi{\begingroup\mn@urlcharsother \@ifnextchar [ {\mn@doi@}
  {\mn@doi@[]}}
\def\mn@doi@[#1]#2{\def\@tempa{#1}\ifx\@tempa\@empty \href
  {http://dx.doi.org/#2} {doi:#2}\else \href {http://dx.doi.org/#2} {#1}\fi
  \endgroup}
\def\mn@eprint#1#2{\mn@eprint@#1:#2::\@nil}
\def\mn@eprint@arXiv#1{\href {http://arxiv.org/abs/#1} {{\tt arXiv:#1}}}
\def\mn@eprint@dblp#1{\href {http://dblp.uni-trier.de/rec/bibtex/#1.xml}
  {dblp:#1}}
\def\mn@eprint@#1:#2:#3:#4\@nil{\def\@tempa {#1}\def\@tempb {#2}\def\@tempc
  {#3}\ifx \@tempc \@empty \let \@tempc \@tempb \let \@tempb \@tempa \fi \ifx
  \@tempb \@empty \def\@tempb {arXiv}\fi \@ifundefined
  {mn@eprint@\@tempb}{\@tempb:\@tempc}{\expandafter \expandafter \csname
  mn@eprint@\@tempb\endcsname \expandafter{\@tempc}}}

\bibitem[\protect\citeauthoryear{{Ali} et~al.,}{{Ali} et~al.}{2015}]{Ali15}
{Ali} Z.~S.,  et~al., 2015, \mn@doi [\apj] {10.1088/0004-637X/809/1/61}, \href
  {http://adsabs.harvard.edu/abs/2015ApJ...809...61A} {809, 61}

\bibitem[\protect\citeauthoryear{{Beardsley} et~al.,}{{Beardsley}
  et~al.}{2016}]{Beardsley16}
{Beardsley} A.~P.,  et~al., 2016, \mn@doi [\apj] {10.3847/1538-4357/833/1/102},
  \href {http://adsabs.harvard.edu/abs/2016ApJ...833..102B} {833, 102}

\bibitem[\protect\citeauthoryear{{Carozzi}}{{Carozzi}}{2015}]{Carozzi15}
{Carozzi} T.~D.,  2015, \mn@doi [\mnras] {10.1093/mnrasl/slv052}, \href
  {http://adsabs.harvard.edu/abs/2015MNRAS.451L...6C} {451, L6}

\bibitem[\protect\citeauthoryear{{Chapman} et~al.,}{{Chapman}
  et~al.}{2013}]{Chapman13}
{Chapman} E.,  et~al., 2013, \mn@doi [\mnras] {10.1093/mnras/sts333}, \href
  {http://adsabs.harvard.edu/abs/2013MNRAS.429..165C} {429, 165}

\bibitem[\protect\citeauthoryear{{Ewall-Wice}, {Dillon}, {Liu}  \&
  {Hewitt}}{{Ewall-Wice} et~al.}{2017}]{Ewall17}
{Ewall-Wice} A.,  {Dillon} J.~S.,  {Liu} A.,   {Hewitt} J.,  2017, \mn@doi
  [\mnras] {10.1093/mnras/stx1221}, \href
  {http://adsabs.harvard.edu/abs/2017MNRAS.470.1849E} {470, 1849}

\bibitem[\protect\citeauthoryear{{Gehlot} et~al.,}{{Gehlot}
  et~al.}{2017}]{Bharat17}
{Gehlot} B.~K.,  et~al., 2017, preprint, \href
  {http://adsabs.harvard.edu/abs/2017arXiv170907727G} {} (\mn@eprint {arXiv}
  {1709.07727})

\bibitem[\protect\citeauthoryear{{Ghosh}, {Mertens}  \& {Koopmans}}{{Ghosh}
  et~al.}{2018}]{Ghosh17}
{Ghosh} A.,  {Mertens} F.~G.,   {Koopmans} L.~V.~E.,  2018, \mn@doi [\mnras]
  {10.1093/mnras/stx2959}, \href
  {http://adsabs.harvard.edu/abs/2018MNRAS.474.4552G} {474, 4552}

\bibitem[\protect\citeauthoryear{{Hazelton}, {Morales}  \&
  {Sullivan}}{{Hazelton} et~al.}{2013}]{Hazelton13}
{Hazelton} B.~J.,  {Morales} M.~F.,   {Sullivan} I.~S.,  2013, \mn@doi [\apj]
  {10.1088/0004-637X/770/2/156}, \href
  {http://adsabs.harvard.edu/abs/2013ApJ...770..156H} {770, 156}

\bibitem[\protect\citeauthoryear{{Jeli{\'c}} et~al.,}{{Jeli{\'c}}
  et~al.}{2008}]{Jelic08}
{Jeli{\'c}} V.,  et~al., 2008, \mn@doi [\mnras]
  {10.1111/j.1365-2966.2008.13634.x}, \href
  {http://adsabs.harvard.edu/abs/2008MNRAS.389.1319J} {389, 1319}

\bibitem[\protect\citeauthoryear{{Koopmans}}{{Koopmans}}{2010}]{Koopmans10}
{Koopmans} L.~V.~E.,  2010, \mn@doi [\apj] {10.1088/0004-637X/718/2/963}, \href
  {http://adsabs.harvard.edu/abs/2010ApJ...718..963K} {718, 963}

\bibitem[\protect\citeauthoryear{{Mertens}, {Ghosh}  \& {Koopmans}}{{Mertens}
  et~al.}{2017}]{Mertens17}
{Mertens} F.~G.,  {Ghosh} A.,   {Koopmans} L.~V.~E.,  2017, preprint, \href
  {http://adsabs.harvard.edu/abs/2017arXiv171110834M} {} (\mn@eprint {arXiv}
  {1711.10834})

\bibitem[\protect\citeauthoryear{{Mesinger}, {Furlanetto}  \& {Cen}}{{Mesinger}
  et~al.}{2011}]{Mesinger11}
{Mesinger} A.,  {Furlanetto} S.,   {Cen} R.,  2011, \mn@doi [\mnras]
  {10.1111/j.1365-2966.2010.17731.x}, \href
  {http://adsabs.harvard.edu/abs/2011MNRAS.411..955M} {411, 955}

\bibitem[\protect\citeauthoryear{{Patil} et~al.,}{{Patil}
  et~al.}{2016}]{Patil16}
{Patil} A.~H.,  et~al., 2016, \mn@doi [\mnras] {10.1093/mnras/stw2277}, \href
  {http://adsabs.harvard.edu/abs/2016MNRAS.463.4317P} {463, 4317}

\bibitem[\protect\citeauthoryear{{Patil} et~al.,}{{Patil}
  et~al.}{2017}]{Patil17}
{Patil} A.~H.,  et~al., 2017, \mn@doi [\apj] {10.3847/1538-4357/aa63e7}, \href
  {http://adsabs.harvard.edu/abs/2017ApJ...838...65P} {838, 65}

\bibitem[\protect\citeauthoryear{Rasmussen \& Williams}{Rasmussen \&
  Williams}{2005}]{Rasmussen05}
Rasmussen C.~E.,  Williams C. K.~I.,  2005, Gaussian Processes for Machine
  Learning (Adaptive Computation and Machine Learning).
The MIT Press

\bibitem[\protect\citeauthoryear{{Thyagarajan} et~al.,}{{Thyagarajan}
  et~al.}{2015}]{Thyagarajan15}
{Thyagarajan} N.,  et~al., 2015, \mn@doi [\apj] {10.1088/0004-637X/804/1/14},
  \href {http://adsabs.harvard.edu/abs/2015ApJ...804...14T} {804, 14}

\bibitem[\protect\citeauthoryear{{Vedantham}, {Udaya Shankar}  \&
  {Subrahmanyan}}{{Vedantham} et~al.}{2012}]{Vedantham12}
{Vedantham} H.,  {Udaya Shankar} N.,   {Subrahmanyan} R.,  2012, \mn@doi [\apj]
  {10.1088/0004-637X/752/2/137}, \href
  {http://adsabs.harvard.edu/abs/2012ApJ...752..137M} {752, 137}

\makeatother
\end{thebibliography}

\end{document}